\documentclass[twocolumn,superscriptaddress,prl,floatfix]{revtex4}
\usepackage[]{graphicx}
\usepackage[usenames,dvipsnames]{color}
\usepackage{soul}
\usepackage{bm}

\begin{document}

\title{Three-Dimensional Electronic Structure of type-II Weyl Semimetal WTe$_2$}

\author{Domenico Di Sante}
\affiliation{Institut f\"{u}r Theoretische Physik und Astrophysik, Universit\"{a}t W\"{u}rzburg, Am Hubland Campus S\"{u}d, W\"{u}rzburg 97074, Germany}\email{domenico.disante@physik.uni-wuerzburg.de}

\author{Pranab Kumar Das}
\affiliation{Istituto Officina dei Materiali (IOM)-CNR, Laboratorio TASC, in Area Science Park, S.S.14, Km 163.5, I-34149 Trieste, Italy}\email{das@iom.cnr.it}
\affiliation{International Centre for Theoretical Physics (ICTP), Strada Costiera 11, I-34100 Trieste, Italy}

\author{C. Bigi}
\affiliation{Dipartimento di Fisica, Universit\'{a} di Milano, Via Celoria 16, I-20133 Milano, Italy}

\author{Z. Erg\"{o}nenc}
\affiliation{Computational Materials Physics, University of Vienna, Sensengasse 8/8, A-1090 Vienna, Austria}

\author{N. G\"urtler}
\affiliation{Computational Materials Physics, University of Vienna, Sensengasse 8/8, A-1090 Vienna, Austria}

\author{J. A. Krieger}
\affiliation{Laboratory for Muon-Spin Spectroscopy, Paul Scherrer Institute, CH-5232 Villigen, Switzerland}
\affiliation{Laboratorium f\"{u}r Festk\"{o}rperphysik, ETH-H\"{o}nggerberg, CH-8093 Z\"{u}rich, Switzerland}

\author{T. Schmitt}
\affiliation{Paul Scherrer Institute, Swiss Light Source, CH-5232 Villigen, Switzerland}

\author{M. N. Ali}
\affiliation{Department of Chemistry, Princeton University, Princeton, New Jersey 08544, USA}

\author{G. Rossi}
\affiliation{Dipartimento di Fisica, Universit\'{a} di Milano, Via Celoria 16, I-20133 Milano, Italy}

\author{R. Thomale}
\affiliation{Institut f\"{u}r Theoretische Physik und Astrophysik, Universit\"{a}t W\"{u}rzburg, Am Hubland Campus S\"{u}d, W\"{u}rzburg 97074, Germany}

\author{C. Franchini}
\affiliation{Computational Materials Physics, University of Vienna, Sensengasse 8/8, A-1090 Vienna, Austria}

\author{S. Picozzi}
\affiliation{Consiglio Nazionale delle Ricerche (CNR-SPIN), Via Vetoio, L'Aquila 67100, Italy}

\author{J. Fujii}
\affiliation{Istituto Officina dei Materiali (IOM)-CNR, Laboratorio TASC, in Area Science Park, S.S.14, Km 163.5, I-34149 Trieste, Italy}

\author{V. N. Strocov}
\affiliation{Paul Scherrer Institute, Swiss Light Source, CH-5232 Villigen, Switzerland}

\author{G. Sangiovanni}
\affiliation{Institut f\"{u}r Theoretische Physik und Astrophysik, Universit\"{a}t W\"{u}rzburg, Am Hubland Campus S\"{u}d, W\"{u}rzburg 97074, Germany}

\author{I. Vobornik}
\affiliation{Istituto Officina dei Materiali (IOM)-CNR, Laboratorio TASC, in Area Science Park, S.S.14, Km 163.5, I-34149 Trieste, Italy}

\author{R. J. Cava}
\affiliation{Department of Chemistry, Princeton University, Princeton, New Jersey 08544, USA}

\author{G. Panaccione}
\affiliation{Istituto Officina dei Materiali (IOM)-CNR, Laboratorio TASC, in Area Science Park, S.S.14, Km 163.5, I-34149 Trieste, Italy}

\date{\today}

\begin{abstract}

By combining bulk sensitive soft-X-ray angular-resolved photoemission
spectroscopy and accurate first-principles calculations we explored the
bulk electronic properties of WTe$_2$, a candidate type-II Weyl
semimetal featuring a large non-saturating magnetoresistance. Despite
the layered geometry suggesting a two-dimensional electronic structure,
we find a three-dimensional electronic dispersion. We report an evident
band dispersion in the reciprocal direction perpendicular to the layers,
implying that electrons can also travel coherently when crossing from
one layer to the other. The measured Fermi surface is characterized by
two well-separated electron and hole pockets at either side of the
$\Gamma$ point, differently from previous more surface sensitive ARPES
experiments that additionally found a significant quasiparticle weight
at the zone center. Moreover, we observe a significant sensitivity of
the bulk electronic structure of WTe$_2$ around the Fermi level to
electronic correlations and renormalizations due to self-energy effects,
previously neglected in first-principles descriptions.

\end{abstract}

\maketitle

{\it Introduction --} The observation of unconventional transport properties in
WTe$_2$ \cite{Ali_WTe2}, such as the large non-saturating
magnetoresistance with values among the highest ever reported, prompted
experiments and theory to address the electronic structure of this
semimetallic transition metal dichalcogenides (TMD)
\cite{Valla_WTe2,WTe2SOC,Arita_WTe2,WTe2QuantumOscillations,
WTe2_3Danisotropy}. WTe$_2$ consists of layers of transition metal (TM) atoms
sandwiched between two layers of chalcogen atoms, similarly to other TMDs
such as MoS$_2$ and MoSe$_2$. Because of the layered structure, TMDs
have commonly been considered as quasi-two-dimensional solids. The
easiness of exfoliation down to a single layer makes them appealing for
nanoscale electronic applications.
WTe$_2$ has also been theoretically described, in a recent paper, as the prototypical
system to host a new topological state of matter called type-II Weyl
semimetal \cite{Weyl-II}. At odds with
standard type-I Weyl semimetals showing a point-like Fermi surface,
type-II Weyl excitations arise at the contact between hole and electron
pockets. Theoretical predictions were immediately
followed by several surface sensitive angle-resolved photoemission
(ARPES) studies claiming evidence of topological Fermi arcs
\cite{WTe2_ARPES1,WTe2_ARPES2,WTe2_ARPES3}.

Our previous investigation by surface sensitive ARPES, spin-resolved
ARPES and DFT calculations, gave clear hints on the non-purely two-dimensional (2D)
electron states of WTe$_2$ and suggested interlayer, i.e. $k$
perpendicular (k$_z$) dispersion and cross-layer compensation of electrons and
holes \cite{PranabNatComm}. However, a direct inspection of the electronic
properties by means of bulk sensitive soft-X-ray ARPES technique, and
more accurate calculations are urgently needed in order to prove the three-dimensional (3D) character
of the bulk electronic structure. By measuring ARPES
with photon energy $h\nu$ in the range 400-800 eV, one probes the
electron states averaging on several layers and therefore reducing the
weight of the surface specific features that otherwise dominate when
excitations energies in the VUV-range are employed.
Furthermore, the increase of photoelectron mean free path in the
soft-X-ray energy range results in a high intrinsic k$_z$ resolution 
of the ARPES experiment \cite{StrokovPRL}, essential to explore 3D effects 
in electronic band structure.

\begin{figure*}[!t]
\centering
\includegraphics[width=\textwidth,angle=0,clip=true]{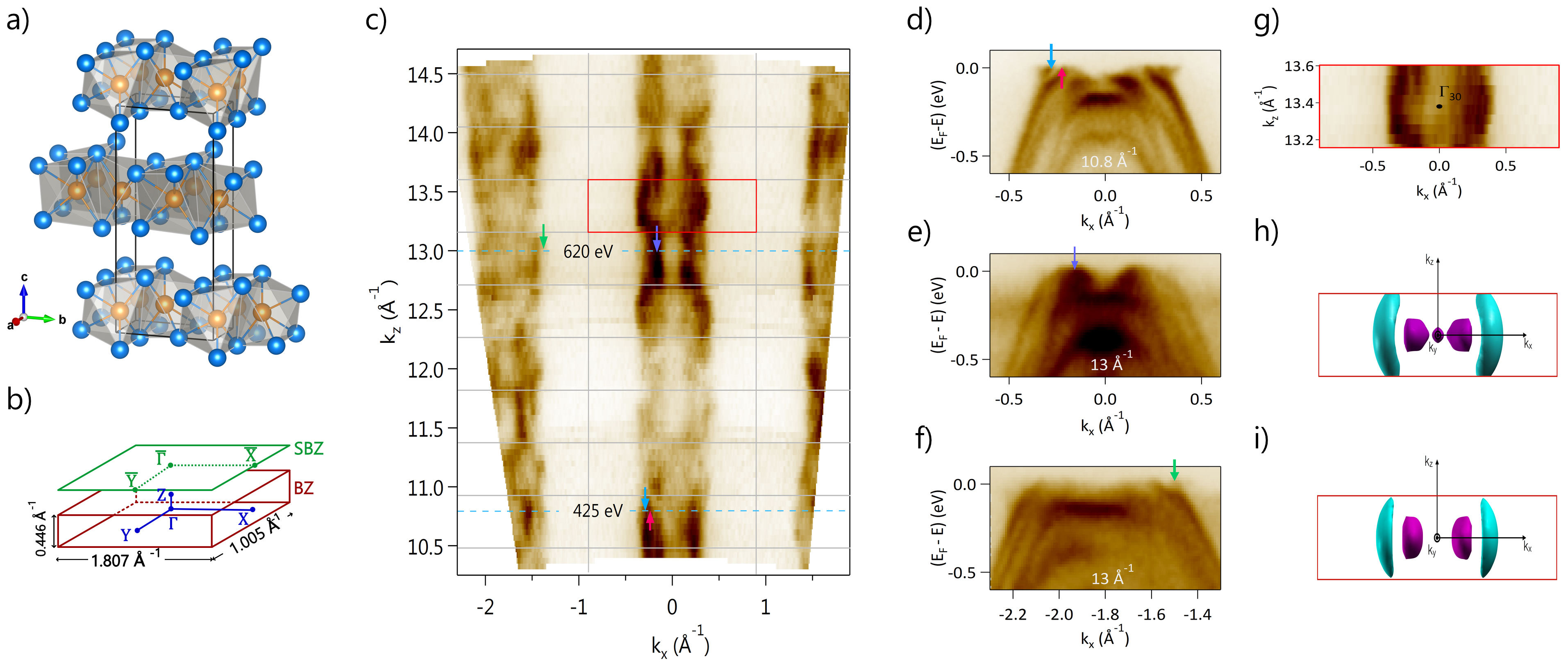}
\caption{(Color online) a) View of the WTe$_2$ crystal structure. W and
Te atoms are shown in orange and blue, respectively. b) Relative bulk
and (001) surface Brillouin zones. c) k$_x$-k$_z$ Fermi surface (k$_y$ = 0)
taken in the 400-800 eV range of excitation energies. The value of the inner
potential V$_0 = -6.5$ eV is used to compute the k$_z$ periodicity over many Brillouin zones (see rectangles),
assumed here to be $2\pi/c\sim 0.45$ \AA$^{-1}$, see below
for a discussion. 
d-f) E vs k band dispersions
along the blue dashed lines in c). Arrows are guides to the eyes to highlight features from hole and electron pockets
in a way consistent with c). g) Zoom of the k$_x$-k$_z$ Fermi surface around
the $\Gamma_{30}$ point (red rectangle in c), and h-i) calculated
Fermi surface within the LDA and LDA+U (U = 2 eV) approximations.}   
\label{fig1}
\end{figure*}

WTe$_2$ displays an unprecedentedly large not-saturating magnetoresistance
even at magnetic fields ${\bf B}$ as high as 60~Tesla \cite{Ali_WTe2}. A
large orbital magnetoresistance is expected in semimetals, so that
WTe$_2$ shares this intriguing feature with bismuth and graphite \cite{BismuthGraphite},
all showing small concentrations of very mobile hole and electron
carriers. Differently from bismuth and graphite, however,
the magnetoresistance in WTe$_2$ exactly follows a ${\bf B}^2$ dependence
typical of an electron-hole compensated semimetal \cite{PippardBOOK}.
Carrier compensation, in turn, is only a necessary condition. It is
equally mandatory that the carrier mobility does not depend on the applied
magnetic field, a feature met by WTe$_2$ \cite{Ali_WTe2} but not for
example by pure bismuth \cite{Bismuth}. The bulk electronic structure of
WTe$_2$ has been so far only investigated by means of transport
measurements \cite{WTe2QuantumOscillations, WTe2_3Danisotropy}. The behavior 
of the resistivity under an external magnetic field is hard to reconcile with 
the picture of a layered solid: when the magnetic
field is applied parallel to the layers, unexpected quantum oscillations
were observed, suggesting that electrons may travel in a coherent way
also across the weakly bonded layers \cite{WTe2_3Danisotropy}. Moreover, recent
spin-ARPES data indicated both in-plane and out-of-plane spin
polarization of the electron states below the Fermi level, deviating
from what expected from spin-orbit interaction (SOI) in a
non-interacting 2D layered system \cite{PranabNatComm}, and the balance between
the hole and electron states was shown to be fully established only if
cooperation of several layers, i.e. bulk 3D character, was included \cite{PranabNatComm}. 
The above direct and indirect evidences indicate that the electronic
structure of WTe$_2$ shows an intrinsic 3D character,
unexpectedly for a TMD. However, up to now, no conclusive spectroscopic
evidence of three-dimensionality was reported.

\begin{figure}[!t]
\centering
\includegraphics[width=\columnwidth,angle=0,clip=true]{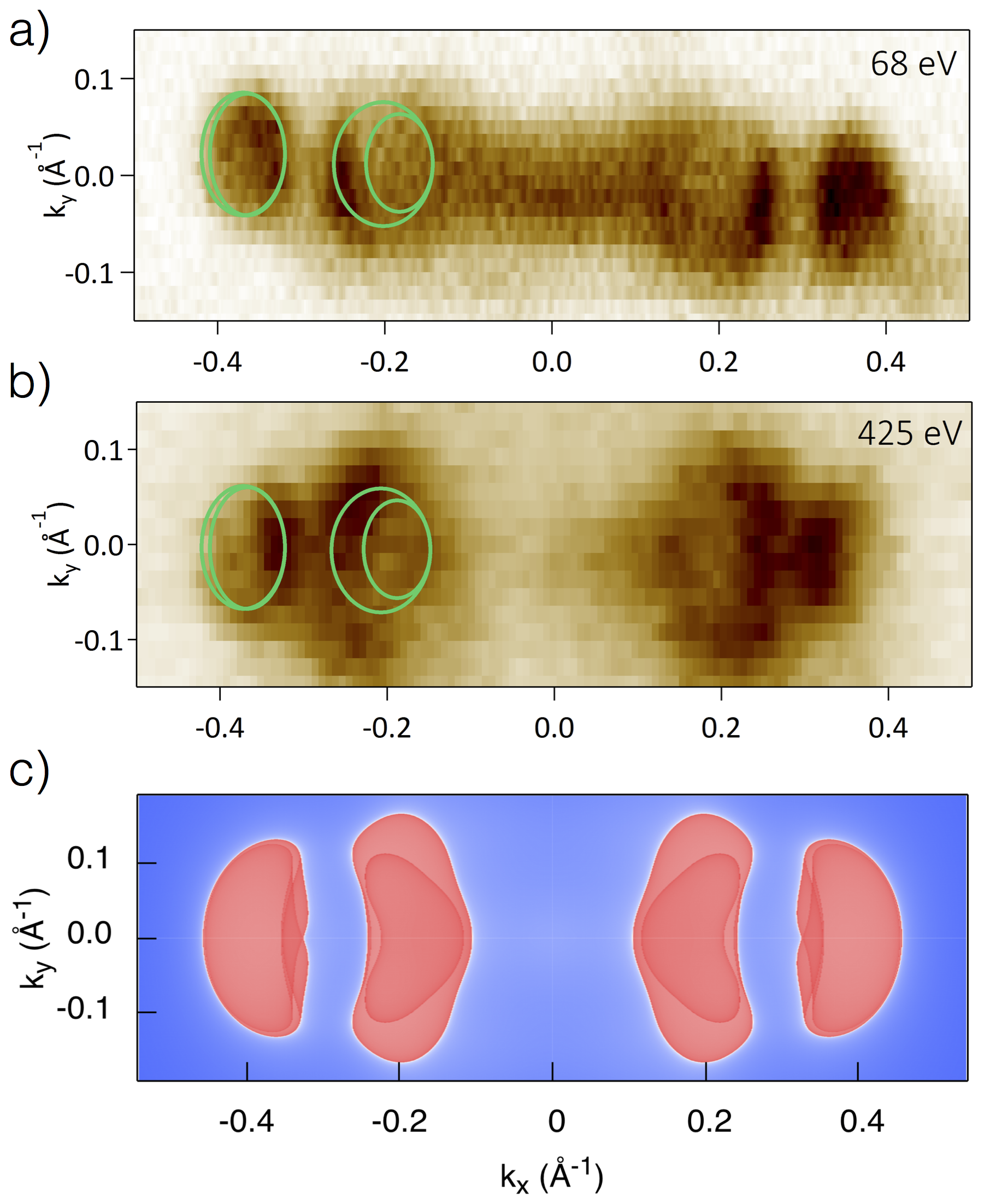}
\caption{(Color online) k$_x$-k$_y$ Fermi surfaces recorded with a) UV
ARPES at $h\nu = 68$ eV and b) soft-X-ray ARPES at $h\nu = 425$ eV, respectively.
Green solid lines are reproduced from Ref. \cite{WTe2Kaminski} corresponding
to the extremal orbits with the areas determined from quantum oscillation
measurements. c) Calculated k$_x$-k$_y$ bulk Fermi surface within the LDA+U
approximation for U = 2 eV.}
\label{fig2}
\end{figure}

{\it Results and Discussion --} In Fig. \ref{fig1}c) we report the k$_z$ evolution
of the Fermi surface spanning about 9 Brillouin zones in the out-of-plane
reciprocal direction and about 3 in-plane Brillouin zones along the W chain direction
k$_x$. Our measurements unambiguously unveil a clear continuous k$_z$ dispersion of the
electronic states at the Fermi level, definitely proving that WTe$_2$ has
a 3D bulk electronic structure despite the layered geometry common to all TMDs \cite{StrokovPRL},
and despite the preferential direction for electronic dispersion given by the zigzag TM chains.
These results provide a spectroscopic validation of the early conclusions based on quantum oscillations
experiments \cite{WTe2_3Danisotropy}.
   
A closer look at the k$_z$ evolution of the
electronic states around k$_x = 0$ would suggest the out-of-plane
periodicity be $\sim 0.9$ \AA$^{-1}$ (two rectangles along k$_z$ in Fig. \ref{fig1}c), as resulting from a unit cell
containing only one WTe$_2$ layer, and contrary to the experimental
structure consisting of two layers, as depicted in Fig. \ref{fig1}a) \cite{WTe2data}.
It is not unusual that the experimentally measured periodicity of the
system is generally different from that of the structural unit cell.
Common examples are given by
nonsymmorphic systems such as CrO$_2$, graphite, and 2H-WSe$_2$ \cite{CrO2,Graphite,WSe2},
for which selection rules restrict the final state symmetry.
WTe$_2$ crystallizes in the nonsymmorphic space group $Pmn2_1$. The nonsymmorphicity 
ensures a symmetry protected band degeneracy at the Brillouin zone point 
$Z = (0,0,\pi/c)$ (see Fig. 1a in the Supp. Mat. \cite{Supp}),
despite the fractional translation invariance along the k$_z$
direction is broken by the non-equivalence of WTe$_2$ monolayers.
To model this double periodicity in band structure calculations,
we have projected the effective band structure of
WTe$_2$ onto the irreducible representations of a Brillouin zone
compatible with the one extracted from our experiments, with restored
k$_z$ fractional translation invariance \cite{VaspUnfolding,WeiKu,Popescu}.
We find indeed that the unfolded band structure, having relatively clean 
unfolded bands around the Fermi level, explains the observed periodicity,
as shown in the Supp. Mat. Fig. 1b \cite{Supp}.

The double k$_z$ extent of the Fermi surface can be further inspected by a closer analysis of
the electronic dispersions \cite{noteSpectra}. At the Fermi level, the
spectrum in Fig. \ref{fig1}d) shows the expected bulk hole and electron
pockets along the k$_x$ direction, as highlighted by light blue and red
arrows, respectively. On the other hand, the spectrum in Fig.
\ref{fig1}e), which misses such features at the Fermi level around k$_x
= 0$, recovers the two pockets at the neighbor in-plane Brillouin zones
around k$_x\sim\pm 1.8$ \AA$^{-1}$ (see Fig. \ref{fig1}f). We, therefore,
conclude that the measured intensities are strongly modified by matrix element
effects, enhancing selectively electron and hole contributions in
subsequent Brillouin zones.

A notable difference between present soft-X-ray data and VUV ones
(see Fig. \ref{fig2}a and Refs.
\cite{PranabNatComm,WTe2PRLFeng}) as well as laser excited ARPES \cite{WTe2Kaminski} is the lack
of any evident spectral intensity at the zone center $\Gamma$.
Nevertheless, the observation of small frequency quantum oscillations
suggests the presence of tiny electron pockets at both side of, and
almost touching at, the $\Gamma$ point
\cite{WTe2QuantumOscillations,WTe2Kaminski}. Our measured Fermi surface
reports no clear evidence of these type of structures up to
a binding energy larger than $\sim 100$ meV from the Fermi level (see Fig. \ref{fig2}b of the main text and Fig. 2 of the Supp. Mat. \cite{Supp}).
In Fig. \ref{fig2}a-b) we also highlight, by means of solid green circles,
the extension of extremal orbits corresponding to given frequencies in
quantum oscillation measurements \cite{WTe2Kaminski}. While such
extensions, approximately of equal size ($\sim 0.025$ \AA$^{-2}$) for
both pockets, nicely fit with our UV ARPES Fermi surface, bulk sensitive
soft-X-ray ARPES shows that the dimension of the hole pocket is much larger
than estimated \cite{CrossSection}. In this respect, our results establish that the
3D dispersion of the Fermi surface is crucial for the
electron-hole compensation that in turn explains the reported giant
magnetoresistance. This is also in line with recent magnetotransport
experiments, supporting the need of three-dimensionality for having such
an extremely large effect \cite{Na_Nanoscale}.

The measured electron pocket is characterized by a bow-like k$_z$
dispersion (Fig. \ref{fig1}g), in agreement with the calculated Fermi
surface based on the local density approximation (LDA) plus an on-site
Hubbard U of 2 eV (compare with cyan areas in Fig. \ref{fig1}i).
However, discrepancies arise when comparing the hole pocket dispersion.
In the measured Fermi surface, hole pockets seem to disperse all over
the k$_z$ extension of the Brillouin zone, while first-principles
calculations give disconnected pockets (purple areas in Fig. \ref{fig1}i).
However, it is worth to note that the calculated k$_x$-k$_y$ Fermi
surface (Fig. \ref{fig2}c)), nicely reproduces the features and the
extensions of the measured Fermi surface. This improves over standard
(i.e. without U) LDA calculations (see Fig. \ref{fig1}h, discussions in Refs.
\cite{Valla_WTe2,WTe2QuantumOscillations,PranabNatComm}, and discussion
below), suggesting that electronic correlation could play a significant
role. In fact, previous theoretical studies have demonstrated that LDA is capable of providing an overall 
good description of the electronic structure of WTe$_2$, especially
the coexistence of electron and hole features and the onset of topological surface states,
but at a quantitative level, significant discrepancies with experiment remain. 
In particular, LDA tends to overestimate the dimensions of the Fermi surface
along the W chains direction (see Fig. \ref{fig1}h), positioning the minimum of the electron pocket at
momentum values larger than ARPES \cite{PranabNatComm}. This, together with the
difficulty in resolving tiny features dispersing around the Fermi level, has limited so far the
understanding of surface and bulk contributions on the Fermi surface. Moreover, since calculations
predict Weyl points to be above the Fermi level, a direct comparison
with experiment requires an accurate treatment of unoccupied states.

\begin{figure}[!t]
\centering
\includegraphics[width=\columnwidth,angle=0,clip=true]{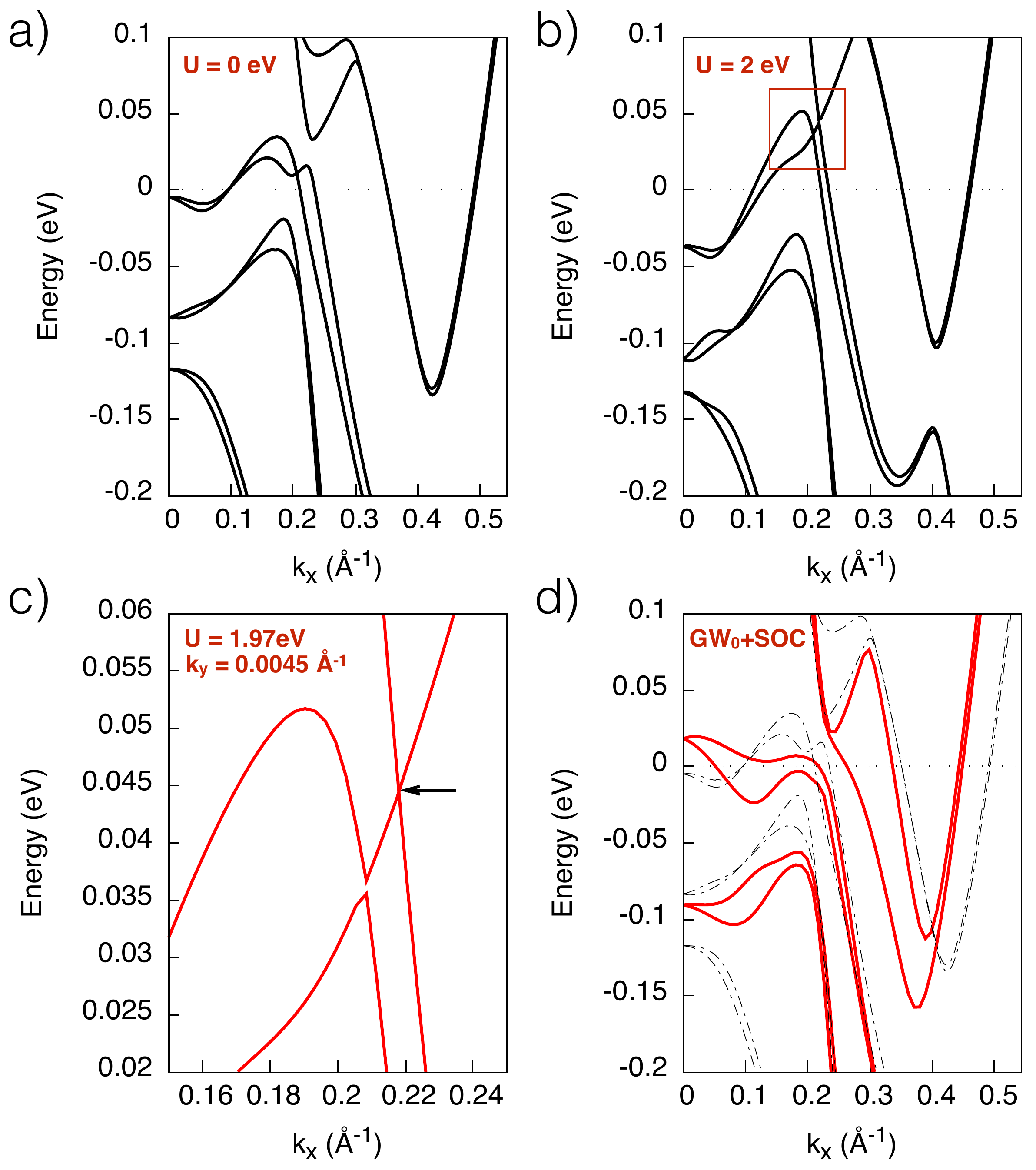}
\caption{(Color online) a-b) LDA+U band structures for U = 0.0 eV and 2.0
eV along the k$_x$ direction, respectively. The red box in panel b)
refers to the zoom area in c) where the 3D linear band
crossing (see arrow) occurs at finite k$_y = \pm 0.0045$ \AA$^{-1}$ for U values smaller than 
the critical U$_c$ = 1.98 eV. 
d) GW band structure (solid red) as compared to LDA (dot-dashed black).}
\label{fig3}
\end{figure}

Despite the fairly delocalized character of 5d W orbitals, the inclusion of a
moderate U in LDA leads to a sizeable modification of the electronic states in the proximity of the
Fermi level, as shown in Fig. \ref{fig3}.
The general trend as
a function of increasing U (see Fig. 4 of the Supplement \cite{Supp} for more values) is the small
shift of the electron pocket toward lower momentum values and the
sizeable modification of the hole pocket. It is interesting to note how
for U = 2 eV, the value that gives a nice comparison between measured
and calculated Fermi surfaces as shown in Fig. \ref{fig2}, the two
pockets almost linearly cross at $\sim 50$ meV above the Fermi level
along the k$_x$ direction. A further increase
of the U (Fig. 4 of Supp. Mat. \cite{Supp}) causes the reopening of the gap between these pockets.
If this correlation driven trend leads to a change of the
topological properties of WTe$_2$ it would deserve a proper investigation.
What we highlight here is that for U values slightly smaller than 2 eV, 
bands linearly cross at finite k$_y$, symmetric with respect the k$_x$ axis, 
giving rise to type-I Weyl points,
as shown in Fig. \ref{fig3}c). Such Weyl points, having opposite chirality,
are connected by Fermi arcs when projected onto the surface Brillouin zone (Fig. 5 of Supp. Mat. \cite{Supp}). At the
critical value U$_c$ = 1.98 eV these Weyl points touch on the k$_x$ axis and
annihilate. The appearance of a type-I Weyl point has
been suggested as the fingerprint of topological transitions in
noncentrosymmetric topological insulators \cite{vanderbilt_prb2014}.

Another different source of band renormalization may be induced by
self-energy effects which we have considered by conducting GW
calculations within a fully relativistic framework. The resulting band
structure, shown in Fig. \ref{fig3}d), displays significant changes with
respect to the underlying DFT (dotted-dashed lines) and DFT+U ones: the
position of the electron pockets shifts closer to the
$\Gamma$-point, in better agreement with experiment, but in contrast
with ARPES and DFT(+U), GW finds a density of empty states at $\Gamma$
right above the Fermi level, which leads to a strong renormalization of
the hole pocket; finally, the quasiparticle bands exhibit a larger
SOI induced splitting, which could indicate a strong
coupling between relativistic effects and electronic correlation, a
novel quantum phenomenon recently observed in other heavy materials
subjected to Lifshitz-type instabilities \cite{KimFranchini}.

{\it Concluding remarks --} In this Letter, we address the bulk
electronic properties of WTe$_2$ by complementary bulk-sensitive
electron spectroscopy and theoretical methods. Since the prediction of
topological surface states in WTe$_2$ owing to a topological nature and
its classification as type-II Weyl semimetal, the spectroscopic study of
the bulk electronic structure of WTe$_2$ was missing.  Our soft-X-ray
ARPES measurements, by means of an 
unprecedentedly high intrinsic definition of k$_z$ and a large range 
of its variation in the Fermi surface mapping,
definitely demonstrate a 3D character of
the electronic states. These results prove that layered materials as
TMDs may host electrons moving from layer to layer in a coherent way,
in agreement with the quantum oscillation transport results
\cite{WTe2_3Danisotropy}. Moreover, our theoretical investigation shed
light on the role of electronic correlations and self-energy
effects on those electronic states, dispersing around the Fermi
level, that play a relevant role in the transport properties of WTe$_2$.

D.D.S., G.S. and R.T. acknowledge the German Research Foundation
(DFG-SFB 1170 Tocotronics), ERC-StG-336012-Thomale-TOPOLECTRICS, NSF
PHY-1125915 and the SuperMUC system at the Leibniz Supercomputing Centre
under the Project-ID pr94vu. 
The soft-X-ray ARPES experiment was carried out at the ADRESS beamline \cite{ADRESS1,ADRESS}
at the Swiss Light Source, Paul Scherrer Institute, Switzerland.
UV-ARPES experiment was performed at APE-IOM beamline at the ELETTRA Sincrotrone Trieste \cite{APE}.
The work at CNR-SPIN and CNR-IOM was
performed within the framework of the nanoscience foundry and fine
analysis (NFFA-MIUR Italy) project. The research in Vienna was supported
by the Austrian Science Fund (Grant No.\ I1490-N19). Compuing time at
the Vienna Scientific Cluster (VSC3) is greatfully acknoledged. The
research at Princeton was supported by the US NSF MRSEC Program Grant
DMR-1420541. J.A.K. was supported by the Swiss National Science
Foundation (SNF-Grant No. 200021-165910).
P.K.D. and D.D.S. contributed equally to this work.

\bibliography{biblio}

\begin{thebibliography}{41}
\expandafter\ifx\csname natexlab\endcsname\relax\def\natexlab#1{#1}\fi
\expandafter\ifx\csname bibnamefont\endcsname\relax
  \def\bibnamefont#1{#1}\fi
\expandafter\ifx\csname bibfnamefont\endcsname\relax
  \def\bibfnamefont#1{#1}\fi
\expandafter\ifx\csname citenamefont\endcsname\relax
  \def\citenamefont#1{#1}\fi
\expandafter\ifx\csname url\endcsname\relax
  \def\url#1{\texttt{#1}}\fi
\expandafter\ifx\csname urlprefix\endcsname\relax\def\urlprefix{URL }\fi
\providecommand{\bibinfo}[2]{#2}
\providecommand{\eprint}[2][]{\url{#2}}

\bibitem[{\citenamefont{Ali et~al.}(2014)\citenamefont{Ali, Xiong, Flynn, Tao,
  Gibson, Schoop, Liang, Haldolaarachchige, Hirschberger, Ong
  et~al.}}]{Ali_WTe2}
\bibinfo{author}{\bibfnamefont{M.~N.} \bibnamefont{Ali}},
  \bibinfo{author}{\bibfnamefont{J.}~\bibnamefont{Xiong}},
  \bibinfo{author}{\bibfnamefont{S.}~\bibnamefont{Flynn}},
  \bibinfo{author}{\bibfnamefont{J.}~\bibnamefont{Tao}},
  \bibinfo{author}{\bibfnamefont{Q.~D.} \bibnamefont{Gibson}},
  \bibinfo{author}{\bibfnamefont{L.~M.} \bibnamefont{Schoop}},
  \bibinfo{author}{\bibfnamefont{T.}~\bibnamefont{Liang}},
  \bibinfo{author}{\bibfnamefont{N.}~\bibnamefont{Haldolaarachchige}},
  \bibinfo{author}{\bibfnamefont{M.}~\bibnamefont{Hirschberger}},
  \bibinfo{author}{\bibfnamefont{N.~P.} \bibnamefont{Ong}},
  \bibnamefont{et~al.}, \bibinfo{journal}{Nature (London)}
  \textbf{\bibinfo{volume}{514}}, \bibinfo{pages}{205} (\bibinfo{year}{2014}).

\bibitem[{\citenamefont{Pletikosi\ifmmode~\acute{c}\else \'{c}\fi{}
  et~al.}(2014)\citenamefont{Pletikosi\ifmmode~\acute{c}\else \'{c}\fi{}, Ali,
  Fedorov, Cava, and Valla}}]{Valla_WTe2}
\bibinfo{author}{\bibfnamefont{I.}~\bibnamefont{Pletikosi\ifmmode~\acute{c}\else
  \'{c}\fi{}}}, \bibinfo{author}{\bibfnamefont{M.~N.} \bibnamefont{Ali}},
  \bibinfo{author}{\bibfnamefont{A.~V.} \bibnamefont{Fedorov}},
  \bibinfo{author}{\bibfnamefont{R.~J.} \bibnamefont{Cava}}, \bibnamefont{and}
  \bibinfo{author}{\bibfnamefont{T.}~\bibnamefont{Valla}},
  \bibinfo{journal}{Phys. Rev. Lett.} \textbf{\bibinfo{volume}{113}},
  \bibinfo{pages}{216601} (\bibinfo{year}{2014}).

\bibitem[{\citenamefont{Jiang et~al.}(2015{\natexlab{a}})\citenamefont{Jiang,
  Tang, Pan, Liu, Niu, Wang, Xu, Yang, Xie, Song et~al.}}]{WTe2SOC}
\bibinfo{author}{\bibfnamefont{J.}~\bibnamefont{Jiang}},
  \bibinfo{author}{\bibfnamefont{F.}~\bibnamefont{Tang}},
  \bibinfo{author}{\bibfnamefont{X.~C.} \bibnamefont{Pan}},
  \bibinfo{author}{\bibfnamefont{H.~M.} \bibnamefont{Liu}},
  \bibinfo{author}{\bibfnamefont{X.~H.} \bibnamefont{Niu}},
  \bibinfo{author}{\bibfnamefont{Y.~X.} \bibnamefont{Wang}},
  \bibinfo{author}{\bibfnamefont{D.~F.} \bibnamefont{Xu}},
  \bibinfo{author}{\bibfnamefont{H.~F.} \bibnamefont{Yang}},
  \bibinfo{author}{\bibfnamefont{B.~P.} \bibnamefont{Xie}},
  \bibinfo{author}{\bibfnamefont{F.~Q.} \bibnamefont{Song}},
  \bibnamefont{et~al.}, \bibinfo{journal}{Phys. Rev. Lett.}
  \textbf{\bibinfo{volume}{115}}, \bibinfo{pages}{166601}
  (\bibinfo{year}{2015}{\natexlab{a}}).

\bibitem[{\citenamefont{Wu et~al.}(2015)\citenamefont{Wu, Jo, Ochi, Huang, Mou,
  Bud'ko, Canfield, Trivedi, Arita, and Kaminski}}]{Arita_WTe2}
\bibinfo{author}{\bibfnamefont{Y.}~\bibnamefont{Wu}},
  \bibinfo{author}{\bibfnamefont{N.~H.} \bibnamefont{Jo}},
  \bibinfo{author}{\bibfnamefont{M.}~\bibnamefont{Ochi}},
  \bibinfo{author}{\bibfnamefont{L.}~\bibnamefont{Huang}},
  \bibinfo{author}{\bibfnamefont{D.}~\bibnamefont{Mou}},
  \bibinfo{author}{\bibfnamefont{S.~L.} \bibnamefont{Bud'ko}},
  \bibinfo{author}{\bibfnamefont{P.~C.} \bibnamefont{Canfield}},
  \bibinfo{author}{\bibfnamefont{N.}~\bibnamefont{Trivedi}},
  \bibinfo{author}{\bibfnamefont{R.}~\bibnamefont{Arita}}, \bibnamefont{and}
  \bibinfo{author}{\bibfnamefont{A.}~\bibnamefont{Kaminski}},
  \bibinfo{journal}{Phys. Rev. Lett.} \textbf{\bibinfo{volume}{115}},
  \bibinfo{pages}{166602} (\bibinfo{year}{2015}).

\bibitem[{\citenamefont{Zhu et~al.}(2015)\citenamefont{Zhu, Lin, Liu, Fauqu\'e,
  Tao, Yang, Shi, and Behnia}}]{WTe2QuantumOscillations}
\bibinfo{author}{\bibfnamefont{Z.}~\bibnamefont{Zhu}},
  \bibinfo{author}{\bibfnamefont{X.}~\bibnamefont{Lin}},
  \bibinfo{author}{\bibfnamefont{J.}~\bibnamefont{Liu}},
  \bibinfo{author}{\bibfnamefont{B.}~\bibnamefont{Fauqu\'e}},
  \bibinfo{author}{\bibfnamefont{Q.}~\bibnamefont{Tao}},
  \bibinfo{author}{\bibfnamefont{C.}~\bibnamefont{Yang}},
  \bibinfo{author}{\bibfnamefont{Y.}~\bibnamefont{Shi}}, \bibnamefont{and}
  \bibinfo{author}{\bibfnamefont{K.}~\bibnamefont{Behnia}},
  \bibinfo{journal}{Phys. Rev. Lett.} \textbf{\bibinfo{volume}{114}},
  \bibinfo{pages}{176601} (\bibinfo{year}{2015}).

\bibitem[{\citenamefont{Thoutam et~al.}(2015)\citenamefont{Thoutam, Wang, Xiao,
  Das, Luican-Mayer, Divan, Crabtree, and Kwok}}]{WTe2_3Danisotropy}
\bibinfo{author}{\bibfnamefont{L.~R.} \bibnamefont{Thoutam}},
  \bibinfo{author}{\bibfnamefont{Y.~L.} \bibnamefont{Wang}},
  \bibinfo{author}{\bibfnamefont{Z.~L.} \bibnamefont{Xiao}},
  \bibinfo{author}{\bibfnamefont{S.}~\bibnamefont{Das}},
  \bibinfo{author}{\bibfnamefont{A.}~\bibnamefont{Luican-Mayer}},
  \bibinfo{author}{\bibfnamefont{R.}~\bibnamefont{Divan}},
  \bibinfo{author}{\bibfnamefont{G.~W.} \bibnamefont{Crabtree}},
  \bibnamefont{and} \bibinfo{author}{\bibfnamefont{W.~K.} \bibnamefont{Kwok}},
  \bibinfo{journal}{Phys. Rev. Lett.} \textbf{\bibinfo{volume}{115}},
  \bibinfo{pages}{046602} (\bibinfo{year}{2015}).

\bibitem[{\citenamefont{Soluyanov et~al.}(2015)\citenamefont{Soluyanov, Gresch,
  Wang, Wu, Troyer, Dai, and Bernevig}}]{Weyl-II}
\bibinfo{author}{\bibfnamefont{A.~A.} \bibnamefont{Soluyanov}},
  \bibinfo{author}{\bibfnamefont{D.}~\bibnamefont{Gresch}},
  \bibinfo{author}{\bibfnamefont{Z.}~\bibnamefont{Wang}},
  \bibinfo{author}{\bibfnamefont{Q.}~\bibnamefont{Wu}},
  \bibinfo{author}{\bibfnamefont{M.}~\bibnamefont{Troyer}},
  \bibinfo{author}{\bibfnamefont{X.}~\bibnamefont{Dai}}, \bibnamefont{and}
  \bibinfo{author}{\bibfnamefont{B.~A.} \bibnamefont{Bernevig}},
  \bibinfo{journal}{Nature (London)} \textbf{\bibinfo{volume}{527}},
  \bibinfo{pages}{495} (\bibinfo{year}{2015}).

\bibitem[{\citenamefont{Bruno et~al.}(2016)\citenamefont{Bruno, Tamai, Wu,
  Cucchi, Barreteau, de~la Torre, McKeown~Walker, Ricc\`o, Wang, Kim
  et~al.}}]{WTe2_ARPES1}
\bibinfo{author}{\bibfnamefont{F.~Y.} \bibnamefont{Bruno}},
  \bibinfo{author}{\bibfnamefont{A.}~\bibnamefont{Tamai}},
  \bibinfo{author}{\bibfnamefont{Q.~S.} \bibnamefont{Wu}},
  \bibinfo{author}{\bibfnamefont{I.}~\bibnamefont{Cucchi}},
  \bibinfo{author}{\bibfnamefont{C.}~\bibnamefont{Barreteau}},
  \bibinfo{author}{\bibfnamefont{A.}~\bibnamefont{de~la Torre}},
  \bibinfo{author}{\bibfnamefont{S.}~\bibnamefont{McKeown~Walker}},
  \bibinfo{author}{\bibfnamefont{S.}~\bibnamefont{Ricc\`o}},
  \bibinfo{author}{\bibfnamefont{Z.}~\bibnamefont{Wang}},
  \bibinfo{author}{\bibfnamefont{T.~K.} \bibnamefont{Kim}},
  \bibnamefont{et~al.}, \bibinfo{journal}{Phys. Rev. B}
  \textbf{\bibinfo{volume}{94}}, \bibinfo{pages}{121112}
  (\bibinfo{year}{2016}).

\bibitem[{\citenamefont{Wang et~al.}(2016)\citenamefont{Wang, Zhang, Huang,
  Nie, Liu, Liang, Zhang, Shen, Liu, Hu et~al.}}]{WTe2_ARPES2}
\bibinfo{author}{\bibfnamefont{C.}~\bibnamefont{Wang}},
  \bibinfo{author}{\bibfnamefont{Y.}~\bibnamefont{Zhang}},
  \bibinfo{author}{\bibfnamefont{J.}~\bibnamefont{Huang}},
  \bibinfo{author}{\bibfnamefont{S.}~\bibnamefont{Nie}},
  \bibinfo{author}{\bibfnamefont{G.}~\bibnamefont{Liu}},
  \bibinfo{author}{\bibfnamefont{A.}~\bibnamefont{Liang}},
  \bibinfo{author}{\bibfnamefont{Y.}~\bibnamefont{Zhang}},
  \bibinfo{author}{\bibfnamefont{B.}~\bibnamefont{Shen}},
  \bibinfo{author}{\bibfnamefont{J.}~\bibnamefont{Liu}},
  \bibinfo{author}{\bibfnamefont{C.}~\bibnamefont{Hu}}, \bibnamefont{et~al.},
  \bibinfo{journal}{Phys. Rev. B} \textbf{\bibinfo{volume}{94}},
  \bibinfo{pages}{241119} (\bibinfo{year}{2016}).

\bibitem[{\citenamefont{Wu et~al.}(2016)\citenamefont{Wu, Mou, Jo, Sun, Huang,
  Bud'ko, Canfield, and Kaminski}}]{WTe2_ARPES3}
\bibinfo{author}{\bibfnamefont{Y.}~\bibnamefont{Wu}},
  \bibinfo{author}{\bibfnamefont{D.}~\bibnamefont{Mou}},
  \bibinfo{author}{\bibfnamefont{N.~H.} \bibnamefont{Jo}},
  \bibinfo{author}{\bibfnamefont{K.}~\bibnamefont{Sun}},
  \bibinfo{author}{\bibfnamefont{L.}~\bibnamefont{Huang}},
  \bibinfo{author}{\bibfnamefont{S.~L.} \bibnamefont{Bud'ko}},
  \bibinfo{author}{\bibfnamefont{P.~C.} \bibnamefont{Canfield}},
  \bibnamefont{and} \bibinfo{author}{\bibfnamefont{A.}~\bibnamefont{Kaminski}},
  \bibinfo{journal}{Phys. Rev. B} \textbf{\bibinfo{volume}{94}},
  \bibinfo{pages}{121113} (\bibinfo{year}{2016}).

\bibitem[{\citenamefont{Das et~al.}(2016)\citenamefont{Das, Di~Sante, Vobornik,
  Fujii, Okuda, Bruyer, Gyenis, Feldman, Tao, Ciancio et~al.}}]{PranabNatComm}
\bibinfo{author}{\bibfnamefont{P.~K.} \bibnamefont{Das}},
  \bibinfo{author}{\bibfnamefont{D.}~\bibnamefont{Di~Sante}},
  \bibinfo{author}{\bibfnamefont{I.}~\bibnamefont{Vobornik}},
  \bibinfo{author}{\bibfnamefont{J.}~\bibnamefont{Fujii}},
  \bibinfo{author}{\bibfnamefont{T.}~\bibnamefont{Okuda}},
  \bibinfo{author}{\bibfnamefont{E.}~\bibnamefont{Bruyer}},
  \bibinfo{author}{\bibfnamefont{A.}~\bibnamefont{Gyenis}},
  \bibinfo{author}{\bibfnamefont{B.~E.} \bibnamefont{Feldman}},
  \bibinfo{author}{\bibfnamefont{J.}~\bibnamefont{Tao}},
  \bibinfo{author}{\bibfnamefont{R.}~\bibnamefont{Ciancio}},
  \bibnamefont{et~al.}, \bibinfo{journal}{Nat. Commun.}
  \textbf{\bibinfo{volume}{7}}, \bibinfo{pages}{10847} (\bibinfo{year}{2016}).

\bibitem[{\citenamefont{Strocov et~al.}(2012)\citenamefont{Strocov, Shi,
  Kobayashi, Monney, Wang, Krempasky, Schmitt, Patthey, Berger, and
  Blaha}}]{StrokovPRL}
\bibinfo{author}{\bibfnamefont{V.~N.} \bibnamefont{Strocov}},
  \bibinfo{author}{\bibfnamefont{M.}~\bibnamefont{Shi}},
  \bibinfo{author}{\bibfnamefont{M.}~\bibnamefont{Kobayashi}},
  \bibinfo{author}{\bibfnamefont{C.}~\bibnamefont{Monney}},
  \bibinfo{author}{\bibfnamefont{X.}~\bibnamefont{Wang}},
  \bibinfo{author}{\bibfnamefont{J.}~\bibnamefont{Krempasky}},
  \bibinfo{author}{\bibfnamefont{T.}~\bibnamefont{Schmitt}},
  \bibinfo{author}{\bibfnamefont{L.}~\bibnamefont{Patthey}},
  \bibinfo{author}{\bibfnamefont{H.}~\bibnamefont{Berger}}, \bibnamefont{and}
  \bibinfo{author}{\bibfnamefont{P.}~\bibnamefont{Blaha}},
  \bibinfo{journal}{Phys. Rev. Lett.} \textbf{\bibinfo{volume}{109}},
  \bibinfo{pages}{086401} (\bibinfo{year}{2012}).

\bibitem[{\citenamefont{Du et~al.}(2005)\citenamefont{Du, Tsai, Maslov, and
  Hebard}}]{BismuthGraphite}
\bibinfo{author}{\bibfnamefont{X.}~\bibnamefont{Du}},
  \bibinfo{author}{\bibfnamefont{S.-W.} \bibnamefont{Tsai}},
  \bibinfo{author}{\bibfnamefont{D.~L.} \bibnamefont{Maslov}},
  \bibnamefont{and} \bibinfo{author}{\bibfnamefont{A.~F.}
  \bibnamefont{Hebard}}, \bibinfo{journal}{Phys. Rev. Lett.}
  \textbf{\bibinfo{volume}{94}}, \bibinfo{pages}{166601}
  (\bibinfo{year}{2005}).

\bibitem[{\citenamefont{Pippard}(1989)}]{PippardBOOK}
\bibinfo{author}{\bibfnamefont{A.~B.} \bibnamefont{Pippard}},
  \emph{\bibinfo{title}{Magnetoresistance in Metals}}
  (\bibinfo{publisher}{Cambridge University Press}, \bibinfo{year}{1989}).

\bibitem[{\citenamefont{Collaudin et~al.}(2015)\citenamefont{Collaudin,
  Fauqu\'e, Fuseya, Kang, and Behnia}}]{Bismuth}
\bibinfo{author}{\bibfnamefont{A.}~\bibnamefont{Collaudin}},
  \bibinfo{author}{\bibfnamefont{B.}~\bibnamefont{Fauqu\'e}},
  \bibinfo{author}{\bibfnamefont{Y.}~\bibnamefont{Fuseya}},
  \bibinfo{author}{\bibfnamefont{W.}~\bibnamefont{Kang}}, \bibnamefont{and}
  \bibinfo{author}{\bibfnamefont{K.}~\bibnamefont{Behnia}},
  \bibinfo{journal}{Phys. Rev. X} \textbf{\bibinfo{volume}{5}},
  \bibinfo{pages}{021022} (\bibinfo{year}{2015}).

\bibitem[{\citenamefont{Wu et~al.}(2017)\citenamefont{Wu, Jo, Mou, Huang,
  Budko, Canfield, and Kaminski}}]{WTe2Kaminski}
\bibinfo{author}{\bibfnamefont{Y.}~\bibnamefont{Wu}},
  \bibinfo{author}{\bibfnamefont{N.~H.} \bibnamefont{Jo}},
  \bibinfo{author}{\bibfnamefont{D.}~\bibnamefont{Mou}},
  \bibinfo{author}{\bibfnamefont{L.}~\bibnamefont{Huang}},
  \bibinfo{author}{\bibfnamefont{S.~L.} \bibnamefont{Budko}},
  \bibinfo{author}{\bibfnamefont{P.~C.} \bibnamefont{Canfield}},
  \bibnamefont{and} \bibinfo{author}{\bibfnamefont{A.}~\bibnamefont{Kaminski}},
  \bibinfo{journal}{ArXiv:1701.06667}  (\bibinfo{year}{2017}).

\bibitem[{\citenamefont{Mar et~al.}(1992)\citenamefont{Mar, Jobic, and
  Ibers}}]{WTe2data}
\bibinfo{author}{\bibfnamefont{A.}~\bibnamefont{Mar}},
  \bibinfo{author}{\bibfnamefont{S.}~\bibnamefont{Jobic}}, \bibnamefont{and}
  \bibinfo{author}{\bibfnamefont{A.}~\bibnamefont{Ibers}}, \bibinfo{journal}{J.
  Am. Chem. Soc.} \textbf{\bibinfo{volume}{114}}, \bibinfo{pages}{8963}
  (\bibinfo{year}{1992}).

\bibitem[{\citenamefont{{Bisti} et~al.}(2016)\citenamefont{{Bisti}, {Rogalev},
  {Karolak}, {Paul}, {Gupta}, {Schmitt}, {G{\"u}ntherodt}, {Sangiovanni},
  {Profeta}, and {Strocov}}}]{CrO2}
\bibinfo{author}{\bibfnamefont{F.}~\bibnamefont{{Bisti}}},
  \bibinfo{author}{\bibfnamefont{V.~A.} \bibnamefont{{Rogalev}}},
  \bibinfo{author}{\bibfnamefont{M.}~\bibnamefont{{Karolak}}},
  \bibinfo{author}{\bibfnamefont{S.}~\bibnamefont{{Paul}}},
  \bibinfo{author}{\bibfnamefont{A.}~\bibnamefont{{Gupta}}},
  \bibinfo{author}{\bibfnamefont{T.}~\bibnamefont{{Schmitt}}},
  \bibinfo{author}{\bibfnamefont{G.}~\bibnamefont{{G{\"u}ntherodt}}},
  \bibinfo{author}{\bibfnamefont{G.}~\bibnamefont{{Sangiovanni}}},
  \bibinfo{author}{\bibfnamefont{G.}~\bibnamefont{{Profeta}}},
  \bibnamefont{and} \bibinfo{author}{\bibfnamefont{V.~N.}
  \bibnamefont{{Strocov}}}, \bibinfo{journal}{ArXiv:1607.01703}
  (\bibinfo{year}{2016}).

\bibitem[{\citenamefont{Pescia et~al.}(1985)\citenamefont{Pescia, Law, Johnson,
  and Hughes}}]{Graphite}
\bibinfo{author}{\bibfnamefont{D.}~\bibnamefont{Pescia}},
  \bibinfo{author}{\bibfnamefont{A.~R.} \bibnamefont{Law}},
  \bibinfo{author}{\bibfnamefont{M.~T.} \bibnamefont{Johnson}},
  \bibnamefont{and} \bibinfo{author}{\bibfnamefont{H.~P.}
  \bibnamefont{Hughes}}, \bibinfo{journal}{Solid State Commun.}
  \textbf{\bibinfo{volume}{56}}, \bibinfo{pages}{809} (\bibinfo{year}{1985}).

\bibitem[{\citenamefont{Finteis et~al.}(1997)\citenamefont{Finteis,
  Hengsberger, Straub, Fauth, Claessen, Auer, Steiner, H\"ufner, Blaha, V\"ogt
  et~al.}}]{WSe2}
\bibinfo{author}{\bibfnamefont{T.}~\bibnamefont{Finteis}},
  \bibinfo{author}{\bibfnamefont{M.}~\bibnamefont{Hengsberger}},
  \bibinfo{author}{\bibfnamefont{T.}~\bibnamefont{Straub}},
  \bibinfo{author}{\bibfnamefont{K.}~\bibnamefont{Fauth}},
  \bibinfo{author}{\bibfnamefont{R.}~\bibnamefont{Claessen}},
  \bibinfo{author}{\bibfnamefont{P.}~\bibnamefont{Auer}},
  \bibinfo{author}{\bibfnamefont{P.}~\bibnamefont{Steiner}},
  \bibinfo{author}{\bibfnamefont{S.}~\bibnamefont{H\"ufner}},
  \bibinfo{author}{\bibfnamefont{P.}~\bibnamefont{Blaha}},
  \bibinfo{author}{\bibfnamefont{M.}~\bibnamefont{V\"ogt}},
  \bibnamefont{et~al.}, \bibinfo{journal}{Phys. Rev. B}
  \textbf{\bibinfo{volume}{55}}, \bibinfo{pages}{10400} (\bibinfo{year}{1997}).

\bibitem[{Sup()}]{Supp}
\bibinfo{note}{See Supplemental Material at http://xxxx.xxxx for experimental
  and computational details, as well as additional ARPES spectra and
  calculations. Supplemental Material includes Refs.
  \cite{VASP1,VASP2,PBE,Dudarev,HedinGW,VASPGW,WANNIER90,VASP2WANNIER90,WTe2data}.}

\bibitem[{\citenamefont{Tomi\'{c} et~al.}(2014)\citenamefont{Tomi\'{c},
  Jeschke, and Valent\'{i}}}]{VaspUnfolding}
\bibinfo{author}{\bibfnamefont{M.}~\bibnamefont{Tomi\'{c}}},
  \bibinfo{author}{\bibfnamefont{H.~O.} \bibnamefont{Jeschke}},
  \bibnamefont{and}
  \bibinfo{author}{\bibfnamefont{R.}~\bibnamefont{Valent\'{i}}},
  \bibinfo{journal}{Phys. Rev. B} \textbf{\bibinfo{volume}{90}},
  \bibinfo{pages}{195121} (\bibinfo{year}{2014}).

\bibitem[{\citenamefont{Ku et~al.}(2010)\citenamefont{Ku, Berlijn, and
  Lee}}]{WeiKu}
\bibinfo{author}{\bibfnamefont{W.}~\bibnamefont{Ku}},
  \bibinfo{author}{\bibfnamefont{T.}~\bibnamefont{Berlijn}}, \bibnamefont{and}
  \bibinfo{author}{\bibfnamefont{C.-C.} \bibnamefont{Lee}},
  \bibinfo{journal}{Phys. Rev. Lett.} \textbf{\bibinfo{volume}{104}},
  \bibinfo{pages}{216401} (\bibinfo{year}{2010}).

\bibitem[{\citenamefont{Popescu and Zunger}(2010)}]{Popescu}
\bibinfo{author}{\bibfnamefont{V.}~\bibnamefont{Popescu}} \bibnamefont{and}
  \bibinfo{author}{\bibfnamefont{A.}~\bibnamefont{Zunger}},
  \bibinfo{journal}{Phys. Rev. Lett.} \textbf{\bibinfo{volume}{104}},
  \bibinfo{pages}{236403} (\bibinfo{year}{2010}).

\bibitem[{not()}]{noteSpectra}
\bibinfo{note}{Spectra in Fig. 1d), e) and f) refer to photon energies $h\nu =
  425$ eV and $620$ eV, such that Fermi surfaces in the first in-plane
  Brillouin zone (k$_x\sim 0$) look different (see k$_z = 10.8$ \AA$^{-1}$ and
  k$_z = 13.0$ \AA$^{-1}$ in Fig. 1c). Spectra for $h\nu = 425$ eV, $570$ eV
  (k$_z = 12.5$ \AA$^{-1}$) and $668$ eV (k$_z = 13.4$ \AA$^{-1}$) are
  similar.}

\bibitem[{\citenamefont{Jiang et~al.}(2015{\natexlab{b}})\citenamefont{Jiang,
  Tang, Pan, Liu, Niu, Wang, Xu, Yang, Xie, Song et~al.}}]{WTe2PRLFeng}
\bibinfo{author}{\bibfnamefont{J.}~\bibnamefont{Jiang}},
  \bibinfo{author}{\bibfnamefont{F.}~\bibnamefont{Tang}},
  \bibinfo{author}{\bibfnamefont{X.~C.} \bibnamefont{Pan}},
  \bibinfo{author}{\bibfnamefont{H.~M.} \bibnamefont{Liu}},
  \bibinfo{author}{\bibfnamefont{X.~H.} \bibnamefont{Niu}},
  \bibinfo{author}{\bibfnamefont{Y.~X.} \bibnamefont{Wang}},
  \bibinfo{author}{\bibfnamefont{D.~F.} \bibnamefont{Xu}},
  \bibinfo{author}{\bibfnamefont{H.~F.} \bibnamefont{Yang}},
  \bibinfo{author}{\bibfnamefont{B.~P.} \bibnamefont{Xie}},
  \bibinfo{author}{\bibfnamefont{F.~Q.} \bibnamefont{Song}},
  \bibnamefont{et~al.}, \bibinfo{journal}{Phys. Rev. Lett.}
  \textbf{\bibinfo{volume}{115}}, \bibinfo{pages}{166601}
  (\bibinfo{year}{2015}{\natexlab{b}}).

\bibitem[{Cro()}]{CrossSection}
\bibinfo{note}{We note here that the observed increment in the size of the hole
  pocket using soft-X-ray is not a consequence of the cross-section variations
  of Te $5p$ and W $5d$ orbitals as both of them are decreasing almost in the
  same manner from UV to the energy ranges used in the present study.}

\bibitem[{\citenamefont{Na et~al.}(2016)\citenamefont{Na, Hoyer, Schoop, Weber,
  Lotsch, Burghard, and Kern}}]{Na_Nanoscale}
\bibinfo{author}{\bibfnamefont{J.}~\bibnamefont{Na}},
  \bibinfo{author}{\bibfnamefont{A.}~\bibnamefont{Hoyer}},
  \bibinfo{author}{\bibfnamefont{L.}~\bibnamefont{Schoop}},
  \bibinfo{author}{\bibfnamefont{D.}~\bibnamefont{Weber}},
  \bibinfo{author}{\bibfnamefont{B.~V.} \bibnamefont{Lotsch}},
  \bibinfo{author}{\bibfnamefont{M.}~\bibnamefont{Burghard}}, \bibnamefont{and}
  \bibinfo{author}{\bibfnamefont{K.}~\bibnamefont{Kern}},
  \bibinfo{journal}{Nanoscale} \textbf{\bibinfo{volume}{8}},
  \bibinfo{pages}{18703} (\bibinfo{year}{2016}).

\bibitem[{\citenamefont{Liu and Vanderbilt}(2014)}]{vanderbilt_prb2014}
\bibinfo{author}{\bibfnamefont{J.}~\bibnamefont{Liu}} \bibnamefont{and}
  \bibinfo{author}{\bibfnamefont{D.}~\bibnamefont{Vanderbilt}},
  \bibinfo{journal}{Phys. Rev. B} \textbf{\bibinfo{volume}{90}},
  \bibinfo{pages}{155316} (\bibinfo{year}{2014}).

\bibitem[{\citenamefont{Kim et~al.}(2016)\citenamefont{Kim, Liu, Erg\"onenc,
  Toschi, Khmelevskyi, and Franchini}}]{KimFranchini}
\bibinfo{author}{\bibfnamefont{B.}~\bibnamefont{Kim}},
  \bibinfo{author}{\bibfnamefont{P.}~\bibnamefont{Liu}},
  \bibinfo{author}{\bibfnamefont{Z.}~\bibnamefont{Erg\"onenc}},
  \bibinfo{author}{\bibfnamefont{A.}~\bibnamefont{Toschi}},
  \bibinfo{author}{\bibfnamefont{S.}~\bibnamefont{Khmelevskyi}},
  \bibnamefont{and}
  \bibinfo{author}{\bibfnamefont{C.}~\bibnamefont{Franchini}},
  \bibinfo{journal}{Phys. Rev. B} \textbf{\bibinfo{volume}{94}},
  \bibinfo{pages}{241113} (\bibinfo{year}{2016}).

\bibitem[{\citenamefont{Strocov et~al.}(2010)\citenamefont{Strocov, Schmitt,
  Flechsig, Schmidt, Imhof, Chen, Raabe, Betemps, Zimoch, Krempasky
  et~al.}}]{ADRESS1}
\bibinfo{author}{\bibfnamefont{V.~N.} \bibnamefont{Strocov}},
  \bibinfo{author}{\bibfnamefont{T.}~\bibnamefont{Schmitt}},
  \bibinfo{author}{\bibfnamefont{U.}~\bibnamefont{Flechsig}},
  \bibinfo{author}{\bibfnamefont{T.}~\bibnamefont{Schmidt}},
  \bibinfo{author}{\bibfnamefont{A.}~\bibnamefont{Imhof}},
  \bibinfo{author}{\bibfnamefont{Q.}~\bibnamefont{Chen}},
  \bibinfo{author}{\bibfnamefont{J.}~\bibnamefont{Raabe}},
  \bibinfo{author}{\bibfnamefont{R.}~\bibnamefont{Betemps}},
  \bibinfo{author}{\bibfnamefont{D.}~\bibnamefont{Zimoch}},
  \bibinfo{author}{\bibfnamefont{J.}~\bibnamefont{Krempasky}},
  \bibnamefont{et~al.}, \bibinfo{journal}{Journal of Synchrotron Radiation}
  \textbf{\bibinfo{volume}{17}}, \bibinfo{pages}{631} (\bibinfo{year}{2010}).

\bibitem[{\citenamefont{Strocov et~al.}(2014)\citenamefont{Strocov, Wang, Shi,
  Kobayashi, Krempasky, Hess, Schmitt, and Patthey}}]{ADRESS}
\bibinfo{author}{\bibfnamefont{V.~N.} \bibnamefont{Strocov}},
  \bibinfo{author}{\bibfnamefont{X.}~\bibnamefont{Wang}},
  \bibinfo{author}{\bibfnamefont{M.}~\bibnamefont{Shi}},
  \bibinfo{author}{\bibfnamefont{M.}~\bibnamefont{Kobayashi}},
  \bibinfo{author}{\bibfnamefont{J.}~\bibnamefont{Krempasky}},
  \bibinfo{author}{\bibfnamefont{C.}~\bibnamefont{Hess}},
  \bibinfo{author}{\bibfnamefont{T.}~\bibnamefont{Schmitt}}, \bibnamefont{and}
  \bibinfo{author}{\bibfnamefont{L.}~\bibnamefont{Patthey}},
  \bibinfo{journal}{Journal of Synchrotron Radiation}
  \textbf{\bibinfo{volume}{21}}, \bibinfo{pages}{32} (\bibinfo{year}{2014}).

\bibitem[{\citenamefont{Panaccione et~al.}(2009)\citenamefont{Panaccione,
  Vobornik, Fujii, Krizmancic, Annese, Giovanelli, Maccherozzi, Salvador,
  Luisa, Benedetti et~al.}}]{APE}
\bibinfo{author}{\bibfnamefont{G.}~\bibnamefont{Panaccione}},
  \bibinfo{author}{\bibfnamefont{I.}~\bibnamefont{Vobornik}},
  \bibinfo{author}{\bibfnamefont{J.}~\bibnamefont{Fujii}},
  \bibinfo{author}{\bibfnamefont{D.}~\bibnamefont{Krizmancic}},
  \bibinfo{author}{\bibfnamefont{E.}~\bibnamefont{Annese}},
  \bibinfo{author}{\bibfnamefont{L.}~\bibnamefont{Giovanelli}},
  \bibinfo{author}{\bibfnamefont{F.}~\bibnamefont{Maccherozzi}},
  \bibinfo{author}{\bibfnamefont{F.}~\bibnamefont{Salvador}},
  \bibinfo{author}{\bibfnamefont{A.~D.} \bibnamefont{Luisa}},
  \bibinfo{author}{\bibfnamefont{D.}~\bibnamefont{Benedetti}},
  \bibnamefont{et~al.}, \bibinfo{journal}{Review of Scientific Instruments}
  \textbf{\bibinfo{volume}{80}}, \bibinfo{pages}{043105}
  (\bibinfo{year}{2009}).

\bibitem[{\citenamefont{Kresse and Furthm\"{u}ller}(1996)}]{VASP1}
\bibinfo{author}{\bibfnamefont{G.}~\bibnamefont{Kresse}} \bibnamefont{and}
  \bibinfo{author}{\bibfnamefont{J.}~\bibnamefont{Furthm\"{u}ller}},
  \bibinfo{journal}{Phys. Rev. B} \textbf{\bibinfo{volume}{54}},
  \bibinfo{pages}{11169} (\bibinfo{year}{1996}).

\bibitem[{\citenamefont{Kresse and Joubert}(1999)}]{VASP2}
\bibinfo{author}{\bibfnamefont{G.}~\bibnamefont{Kresse}} \bibnamefont{and}
  \bibinfo{author}{\bibfnamefont{D.}~\bibnamefont{Joubert}},
  \bibinfo{journal}{Phys. Rev. B} \textbf{\bibinfo{volume}{59}},
  \bibinfo{pages}{1758} (\bibinfo{year}{1999}).

\bibitem[{\citenamefont{Perdew et~al.}(1996)\citenamefont{Perdew, Burke, and
  Ernzerhof}}]{PBE}
\bibinfo{author}{\bibfnamefont{J.~P.} \bibnamefont{Perdew}},
  \bibinfo{author}{\bibfnamefont{K.}~\bibnamefont{Burke}}, \bibnamefont{and}
  \bibinfo{author}{\bibfnamefont{M.}~\bibnamefont{Ernzerhof}},
  \bibinfo{journal}{Phys. Rev. Lett.} \textbf{\bibinfo{volume}{77}},
  \bibinfo{pages}{3865} (\bibinfo{year}{1996}).

\bibitem[{\citenamefont{Dudarev et~al.}(1998)\citenamefont{Dudarev, Botton,
  Savrasov, Humphreys, and Sutton}}]{Dudarev}
\bibinfo{author}{\bibfnamefont{S.~L.} \bibnamefont{Dudarev}},
  \bibinfo{author}{\bibfnamefont{G.~A.} \bibnamefont{Botton}},
  \bibinfo{author}{\bibfnamefont{S.~Y.} \bibnamefont{Savrasov}},
  \bibinfo{author}{\bibfnamefont{C.~J.} \bibnamefont{Humphreys}},
  \bibnamefont{and} \bibinfo{author}{\bibfnamefont{A.~P.}
  \bibnamefont{Sutton}}, \bibinfo{journal}{Phys. Rev. B}
  \textbf{\bibinfo{volume}{57}}, \bibinfo{pages}{1505} (\bibinfo{year}{1998}).

\bibitem[{\citenamefont{Hedin}(1965)}]{HedinGW}
\bibinfo{author}{\bibfnamefont{L.}~\bibnamefont{Hedin}},
  \bibinfo{journal}{Phys. Rev.} \textbf{\bibinfo{volume}{139}},
  \bibinfo{pages}{A796} (\bibinfo{year}{1965}).

\bibitem[{\citenamefont{Shishkin and Kresse}(2006)}]{VASPGW}
\bibinfo{author}{\bibfnamefont{M.}~\bibnamefont{Shishkin}} \bibnamefont{and}
  \bibinfo{author}{\bibfnamefont{G.}~\bibnamefont{Kresse}},
  \bibinfo{journal}{Phys. Rev. B} \textbf{\bibinfo{volume}{74}},
  \bibinfo{pages}{035101} (\bibinfo{year}{2006}).

\bibitem[{\citenamefont{Mostofi et~al.}(2008)\citenamefont{Mostofi, Yates, Lee,
  Souza, Vanderbilt, and Marzari}}]{WANNIER90}
\bibinfo{author}{\bibfnamefont{A.~A.} \bibnamefont{Mostofi}},
  \bibinfo{author}{\bibfnamefont{J.~R.} \bibnamefont{Yates}},
  \bibinfo{author}{\bibfnamefont{Y.-S.} \bibnamefont{Lee}},
  \bibinfo{author}{\bibfnamefont{I.}~\bibnamefont{Souza}},
  \bibinfo{author}{\bibfnamefont{D.}~\bibnamefont{Vanderbilt}},
  \bibnamefont{and} \bibinfo{author}{\bibfnamefont{N.}~\bibnamefont{Marzari}},
  \bibinfo{journal}{Comput. Phys. Commun.} \textbf{\bibinfo{volume}{178}},
  \bibinfo{pages}{685} (\bibinfo{year}{2008}).

\bibitem[{\citenamefont{Franchini et~al.}(2012)\citenamefont{Franchini,
  Kov\'{a}\v{c}ik, Marsman, Sathyanarayana~Murthy, He, Ederer, and
  Kresse}}]{VASP2WANNIER90}
\bibinfo{author}{\bibfnamefont{C.}~\bibnamefont{Franchini}},
  \bibinfo{author}{\bibfnamefont{R.}~\bibnamefont{Kov\'{a}\v{c}ik}},
  \bibinfo{author}{\bibfnamefont{M.}~\bibnamefont{Marsman}},
  \bibinfo{author}{\bibfnamefont{S.}~\bibnamefont{Sathyanarayana~Murthy}},
  \bibinfo{author}{\bibfnamefont{J.}~\bibnamefont{He}},
  \bibinfo{author}{\bibfnamefont{C.}~\bibnamefont{Ederer}}, \bibnamefont{and}
  \bibinfo{author}{\bibfnamefont{G.}~\bibnamefont{Kresse}},
  \bibinfo{journal}{J. Phys. Condens. Matter.} \textbf{\bibinfo{volume}{24}},
  \bibinfo{pages}{235602} (\bibinfo{year}{2012}).

\end{thebibliography}

\end{document}